# Cross-Tokamak Deployment Study of Plasma Disruption Predictors Based on Convolutional Autoencoder

X.K. Ai[1], W. Zheng[1]*, M. Zhang[1], Y.H. Ding[1], D.L. Chen[2], Z.Y. Chen[1], C.S. Shen[1], B.H. Guo[2], N.C. Wang[1], Z.J. Yang[1], Z.P. Chen[1], Y. Pan[1], B. Shen[2], B.J. Xiao[2], and J-TEXT team[1]

[1] State Key Laboratory of Advanced Electromagnetic Technology, International Joint Research Laboratory of Magnetic Confinement Fusion and Plasma Physics, School of Electrical and Electronic Engineering, Huazhong University of Science and Technology, Wuhan 430074, China

[2] Institute of Plasma Physics, Chinese Academy of Sciences, Hefei 230031, China

Corresponding author: W. Zheng

E-mail: zhengwei@hust.edu.cn

**Fax:** +86-27-87793060

## Abstract

In the initial stages of operation for future tokamak, facing limited data availability, deploying data-driven disruption predictors requires optimal performance with minimal use of new device data. This paper studies the issue of data utilization in data-driven disruption predictor during cross tokamak deployment. Current predictors primarily employ supervised learning methods and require a large number of disruption and non-disruption shots for training. However, the scarcity and high cost of obtaining disruption shots for future tokamaks result in imbalanced training datasets, reducing the performance of supervised learning predictors. To solve this problem, we propose the Enhanced Convolutional Autoencoder Anomaly Detection (E-CAAD) predictor. E-CAAD can be only trained by normal samples from non-disruption shots and can also be trained by disruption precursor samples when disruption shots occur. This model not only overcomes the sample imbalance in supervised learning predictors, but also overcomes the inefficient dataset utilization faced by traditional anomaly detection predictors that cannot use disruption precursor samples for training, making it more suitable for the unpredictable datasets of future tokamaks. Compared to traditional anomaly detection predictor, the E-CAAD predictor performs better in disruption prediction and is deployed faster on new devices. Additionally, we explore strategies to accelerate deployment of E-CAAD predictor on the new device by using data from existing devices. Two deployment strategies are presented: mixing data from existing devices and fine-tuning the predictor trained on existing devices. Our comparisons indicate that the data from existing device can accelerate the deployment of predictor on new device. Notably, the fine-tuning strategy yields the fastest deployment on new device among the designed strategies.

**Keywords:** disruption prediction, deep learning, anomaly detection, cross-tokamak, transfer learning.

# 1 Introduction

Plasma disruption, a ubiquitous occurrence in tokamaks, entails the sudden loss of plasma thermal energy and current quench. This phenomenon poses significant challenges for future devices, particularly in larger tokamaks like ITER. Here, the electromagnetic force and thermal load will become even greater, leading to more severe damage and economic losses caused by disruptions[1]-[3]. Plasma instabilities are commonly observed before the occurrence of disruptions, providing an opportunity for disruption prediction. Detection of disruption precursors can trigger the activation of the disruption mitigation system (DMS) to reduce damage to the tokamak[4]. However, due to the incomplete understanding of plasma disruption in tokamaks, developing a reliable physics-driven predictor remains challenging. Physics-driven predictors, such as those using the locked mode signal, have been tested on JET[5] but exhibit inadequate performance for future high-power tokamaks like ITER.

Data-driven disruption predictors primarily use supervised learning methods. Various supervised learning algorithms, including neural networks[6]-[10], support vector machines[11]-[13], and random forests[14],[15], have been employed to develop disruption predictors in existing devices such as J-TEXT[16],[17], EAST[18],[19], HL-2A[20],[21], JET[22]-[24], ASDEX-U[25]-[27], DIII-D[28],[29], and JT-60U[30],[31]. These predictors have shown good performance, some predictors can achieve a success rate of over 90% and a false alarm rate of less than 10%. However, for future high-power reactors like ITER, obtaining an ample quantity and variety of disruption precursor data without damaging the device is not feasible. Consequently, the dataset becomes highly imbalanced, primarily consisting of non-disruption shot data. This severe data imbalance leads to reduced or even ineffective performance of supervised learning predictors[32],[33], hindering their deployment on future high-performance devices.

The deployment of disruption predictors across different tokamaks presents additional challenges due to differences in structure, operating parameters, and control systems[34]. The performance of high-performance pre-trained models from existing devices is unsatisfactory when directly deployed on new devices[35]. Some preliminary studies have explored solutions to this issue and found that in the absence of high performance (HP) data from the target devices, the best predictivity of the HP regime for the target machine can be achieved by combining low performance (LP) data from the target with HP data from other machines[36]. But the future tokamak, such as ITER, will be operated from the first shot, and the data set is scarce in the early operation period, there could be not enough data to build a good enough model. Other studies have been conducted to explore strategies for establishing predictors from scratch, aiming to solve the data scarce problem[37]-[39]. In those study, the supervised learning predictors start training with few shots as the initial training set, and then shots are added in chronological order to the training set to retrain predictor. This retraining process updates the predictor to adapt to changes in the operating environment of new device, which is also known as adaptive learning. And in Dormido-Canto's article[37], balanced and unbalanced datasets are used to develop predictors from scratch, The results also show that unbalanced datasets have a negative impact on the deployment of supervised learning predictors on new devices.

In this paper, we propose a convolutional autoencoder anomaly detection (CAAD) predictor, which is trained with only normal samples of non-disruption shots. The CAAD predictor overcomes the sample imbalance issue of supervised learning predictors. However, in the early operation of future tokamaks, both non-disruption and disruption shots are scarce. The traditional anomaly detection model is trained only on non-disruption shots, resulting in a waste of disruption precursor samples and inefficient dataset utilization. To address this issue, we improved the loss function of the CAAD model to obtain the Enhanced Convolutional Autoencoder Anomaly Detection (E-CAAD) predictor. The E-CAAD predictor can be trained using only normal samples from non-disruption shots and can also be trained using disruption precursor samples when available. This approach not only overcomes the sample imbalance issue of supervised learning predictors but also overcomes the inefficient dataset utilization of traditional anomaly detection predictors. These attributes make the E-CAAD predictor well-suited for the unpredictable datasets of future tokamaks. Additionally, the convolutional autoencoder neural network devised in this study has an intrinsic ability to automatically extract features from raw signals. Given the incomplete understanding of plasma disruption theory, establishing reliable manual feature extraction algorithms is a challenging task. Inaccurate manual extraction methods can result in the loss of crucial information within the samples. The convolutional autoencoder neural network effectively addresses this challenge by incorporating multiple layers of convolutional units. This design allows the model to train with the raw slices of the diagnostic signals, ensuring that all pertinent physical information is captured from the input signals. The model autonomously extracts features from the raw input signals during training, effectively learning the physical data conveyed within the diagnostic signals.

When deploying CAAD and E-CAAD predictors on J-TEXT and EAST, the results show good prediction performance on both devices, with the E-CAAD predictor outperforming the CAAD predictor. When considering J-TEXT and EAST as new devices for deployment experiments starting from scratch, it is observed that the E-CAAD predictor deploys faster on the new device compared to the CAAD predictor. Finally, in cross- tokamak deployment experiment, J-TEXT is regarded as the existing device, and EAST is regarded as the new device. We explore strategies to accelerate model deployment on the new device by leveraging data from existing devices, with a focus on achieving good performance using minimal new device data during the initial operation stages. Two cross-tokamak deployment strategies are presented: mixing data from existing devices and fine-tuning the predictor trained on existing devices. Our comparisons indicate that the data from existing device can accelerate the deployment of predictor on new device. Notably, the fine-tuning strategy yields the fastest deployment on new device among the proposed strategies.

This paper is organized as follows: Section 2 introduces the principle of the CAAD and E-CAAD predictors. Section 3 presents the performance of these predictors, including: prediction performance on existing device and deployment from scratch on new device. Section 4 proposes some cross-tokamak deployment strategies and compares them. Section 5 summarizes the content of the article.

## 2 Design of Convolutional Autoencoder Disruption Predictor

The data-driven disruption predictor monitors the state of the plasma based on diagnostic signals collected by sensors on the tokamak. In this paper, a convolutional autoencoder neural network is designed for disruption prediction according to the characteristics of the slices of selected diagnostic data. In Section 2.1, the diagnostic selection and slicing method in the disruption prediction of this paper are introduced. The network structure of the convolutional autoencoder and the CAAD predictor and E-CAAD predictor for disruption prediction are introduced in Section 2.2.

### 2.1 Diagnostic Selection and Slicing

When selecting diagnostics for disruption prediction, it is essential to choose diagnostic signals related to disruption and ensure that those signals can be collected in real-time. Since this study involves cross-tokamak deployment of the predictor between J-TEXT and EAST, the selection of diagnostic signals must also consider signals that exist on both devices. Table 1 presents the selection and slicing of diagnostic signals. XUV array and Soft-X array measure radiation power and temperature of plasma profile, which also provide relevant information about impurities in the plasma. The data of the CIII array and Ha array are unavailable from EAST team in the conducted experiments, leading to their exclusion from the diagnostics. For vertical displacement event (VDE), the centroid position diagnostics (dx, dy) of the plasma current are selected. Due to the prevalent use of the limiter configuration in J-TEXT discharges, diagnostics related to the plasma elongation ratio were excluded. For the density limit disruptions, the plasma density array is selected. Additionally, the saddle loop array and Mirnov array are selected to detect magnetohydrodynamics (MHD) instabilities such as locked modes and n=1 tearing modes. For non-classifiable disruptions, the macro signals such as plasma current, toroidal magnetic field, and loop voltage are selected. These selected diagnostics are mainly divided into two types: single-channel diagnostics (e.g., Ip, Bt) and multi-channel diagnostics (e.g., Mirnov array, ne array). Single-channel diagnostic only has one time series and only includes the temporal feature of the diagnostic. Multi-channel diagnostic refers to the collection of multiple time series by an observation array composed of multiple sensors, including the temporal and spatial features of diagnostic. For the same multi-channel diagnostic on J-TEXT and EAST, the observation fields of the selected channels are similar in space.

Table 1 Descriptions and symbols of all the features

| Signal | Diagnostic | Channel | Sampling rate [kHz] | Time window [ms] |
|---|---|---|---|---|
| Plasma current | Ip | 1 | 1 | 10 |
| Toroidal magnetic field | Bt | 1 | 1 | 10 |
| Plasma horizontal displacement | dx | 1 | 1 | 10 |
| Plasma vertical displacement | dy | 1 | 1 | 10 |
| Loop voltage | $V_{loop}$ | 1 | 1 | 10 |
| Plasma density | ne array | 11 | 1 | 10 |
| Soft X-ray | SXR array | 30 | 1 | 10 |
| Radiation | XUV array | 32 | 1 | 10 |

|  | Saddle loop detectors | 4 | 10 | 4 |
| --- | --- | --- | --- | --- |
| MHD instabilities related | Mirnov toroidal array | 8 | 60 | 1 |
|  | Mirnov poloidal array | 24 | 60 | 1 |

In our disruption prediction experiment, the slices of diagnostics from J-TEXT and EAST are used by predictors. The input samples for the predictor has a sampling rate of 1 kHz. Each sample consists of slices from the diagnostic signals, where the length of each slice is determined by multiplying the corresponding diagnostic signal's sampling rate by the chosen time window size. Different diagnostics have different window sizes. The last point of each diagnostic signal slice represents the same time moment, representing the time of the sample. To reduce the computational load on the samples, down sampling is applied to each input diagnostic signal while preserving signal integrity based on the Nyquist-Shannon sampling theorem[42]. For example, by performing spectral analysis on Mirnov signals from J-TEXT and EAST, we observe significant spectral differences between the disruption precursor period and normal operation period within 30 kHz. Consequently, we set the sampling rate for the Mirnov signal at 60 kHz. Similar strategies are employed to determine the sampling rates for other diagnostics, with a minimum sampling rate set at 1 kHz. For diagnostics without oscillation features, we set the time window size at 10 ms. For diagnostics with oscillation features, the time window length is set to be longer than one period of the main harmonic oscillation.

## 2.2 Convolutional Autoencoder Disruption Predictor

The autoencoder is a neural network-based deep learning algorithm that performs compression and reconstruction of input samples[40],[41]. When applied to anomaly detection, the model is trained only with normal samples. The similarity between input samples and their reconstructed counterparts is quantified using the mean squared error (MSE). Normal samples result in well reconstructions with small MSE, while abnormal samples yield poor reconstructions with a large MSE. According to the characteristics of the selected diagnostics in Section 2.1, we propose a neural network based on convolutional autoencoder, as depicted in Figure 1. The network structure comprises three main components: convolution feature extraction, feature autoencoder, and transposed convolution reconstruction. The convolution feature extraction module performs feature extraction on the slices of the selected raw signals for each diagnostic. The extracted features from each diagnostic are flattened and concatenated to form a one-dimensional vector that encompasses all diagnostic features. This one-dimensional vector is then inputted into the feature autoencoder module, which consists of five fully connected layers, for compression and reconstruction. The resulting reconstructed one-dimensional vector is subsequently split to obtain the reconstructed features for each diagnostic. The transposed convolution reconstruction module then restores the reconstructed features of each diagnostic to their corresponding reconstructed input slices. Since the input slice shapes may differ for various diagnostics, the structures of the convolutional feature extraction modules and transposed convolution reconstruction modules may not necessarily be the same. Diagnostics can be broadly classified into two types: single-channel diagnostics and multi-channel diagnostics. For single-channel diagnostic slices, the feature extraction module employs one-dimensional convolution to extract temporal features, while the reconstruction module uses one-dimensional transposed convolution to restore the slices. For multi-channel diagnostic slices, the feature extraction module employs two-dimensional convolution to extract combined temporal and spatial features, and the reconstruction module employs two-dimensional transposed convolution to restore the slices. Notably, for the same diagnostic, the structures of the convolutional feature extraction module and transposed convolution reconstruction module are mirror images of each other. The convolutional autoencoder network compresses and reconstructs the input samples. The dimension of the hidden layer in this network is 200, which is much smaller than the input sample. Therefore, the network has the ability to automatically extract features from the original signal. Overall, the reconstruction error ($RCE$) of convolutional autoencoder is the average of mean squared errors of all selected diagnostic, which is:

$$RCE = \frac{1}{n_{diag}}(MSE_{ip} + MSE_{bt} + \cdots + MSE_{mirnov}) \qquad (2\text{-}1)$$

Where $n_{diag}$ is the number of selected diagnostics.

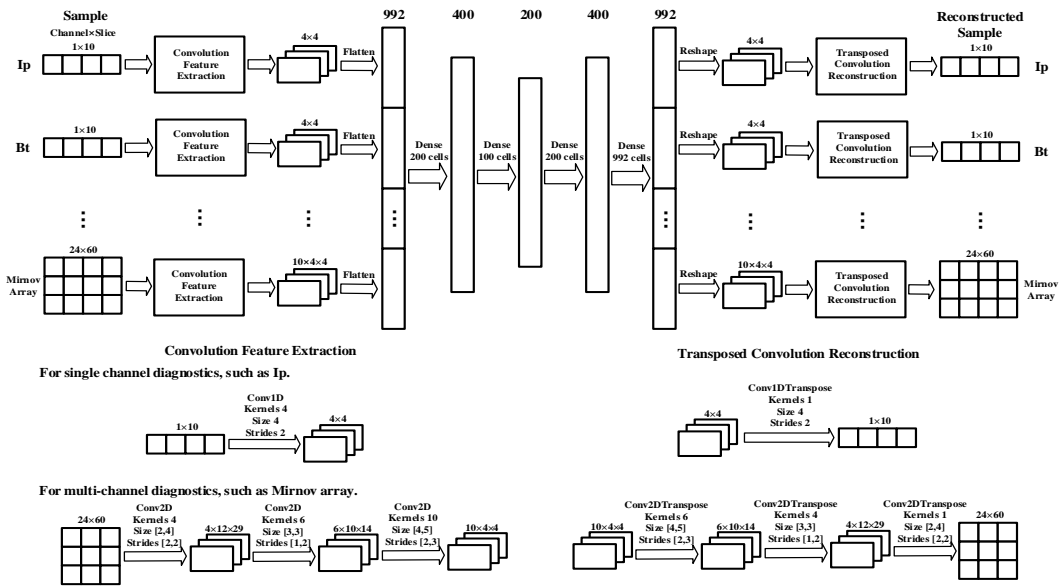

Figure 1 Network structure of convolutional autoencoder

## Anomaly Detection Predictor

When utilizing the convolutional autoencoder for anomaly detection, the loss function of neural network is:

$$loss = \frac{1}{n}\sum_{i=1}^{n} RCE_i \qquad (2\text{-}2)$$

where $n$ is the number of training samples.

The model only uses the normal samples of non-disruption shot for training, aiming to minimize the reconstruction error ($RCE$) specifically for these samples. During the inference process, for non-disruption samples, the returned $RCE$s are small; For disruption precursor samples, the model has not learned the corresponding information, and the returned $RCE$s are relatively large. The alarm threshold of $RCE$ is set for disruption prediction. When the $RCE$ of the real-time sample exceeds the threshold, the disruption alarm is issued.

### Enhanced Anomaly Detection Predictor

When deploying on new device, non-disruption and disruption shots are scarce during the early operation stage, the anomaly detection models cannot be trained with disruption precursor samples, resulting in inefficient dataset utilization. To solve this problem, we improve the loss function of the convolutional autoencoder. We improve the loss function of the convolutional autoencoder so that the predictor can be trained with the disruption precursor samples when the disruption shot appears. The improved loss function is:

$$loss = \frac{1}{n}\sum_{i=1}^{n}[y \cdot RCE_i^y + (1-y) \cdot \alpha \cdot RCE_i^{y-1}] \tag{2-3}$$

where $y$ is the label of input sample. The non-disruption sample is labeled as 1, the disruption precursor sample is labeled as 0. $\alpha$ is a hyperparameter, which adjusts the weight of the model to the disruption precursor samples. $n$ is the number of training samples.

During the training process, the model is optimized to minimize $RCE$ of non-disruption samples and $RCE^{-1}$ of disruption precursor samples, which means the model maximizes the $RCE$ of disruption precursor samples. Therefore, the model can be trained with disruption precursor samples when the disruption occurs, and can still be trained with only normal samples when there are no disruption shots in the training set. This attribute makes the E-CAAD predictor well-suited for the unpredictable datasets of future tokamaks. During the inference process, for non-disruption samples, the returned $RCE$s are relatively small; for disruption precursor samples, the returned $RCE$s are relatively large. The alarm threshold of $RCE$ is set for disruption prediction. When the $RCE$ of the real-time sample exceeds the threshold, the disruption alarm is issued.

## 3  Performance Comparison

Disruption prediction experiments are conducted on the CAAD and E-CAAD predictor using the data from J-TEXT and EAST tokamak devices. The J-TEXT dataset comprised discharges within the shot range of 1045962 to 1066648, spanning from 2017 to 2019. During this period the typical discharge of the J-TEXT in the limiter configuration is done with a plasma current $I_p$ of around 200 kA, a toroidal field $B_t$ of around 2.0 T, a pulse length of around 800 ms, plasma densities $n_e$ of around 1–7 $\times 10^{19}$ m$^{-3}$, and an electron temperature $T_e$ of about 1 keV. And EAST dataset included discharges within the shot range of 54000 to 97000, spanning from 2015 to 2019. During this period the typical discharge of the EAST tokamak in the divertor configuration is done with a plasma current $I_p$ of around 450 kA, a toroidal field $B_t$ of around 1.5 T, a pulse length of around 10 s, and a $\beta_N$ of around 2.1. To ensure the reliability of the experiments, intentional disruption shots are excluded from the training and analysis. For both datasets, shots with complete diagnostic signals are randomly selected from the remaining shots to perform the experiments on each device. The split of the dataset is shown in Table 2. The comparisons of the disruption prediction performance of CAAD and E-CAAD is conducted by using this dataset. There are two types of comparative experiments: one is to compare the performance of predictors trained by ample data from J-TEXT and EAST in Section 3.1. The other is to regard J-TEXT and EAST as new devices and conducts deployment experiments from scratch, comparing the deployment speed of predictors on new devices in Section 3.2.

Table 2 Dataset split on J-TEXT and EAST device

|  | Non-disruption [J-TEXT] | Disruption [J-TEXT] | Non-disruption [EAST] | Disruption [EAST] |
|---|---|---|---|---|
| No. training shots | 300 | 225 | 400 | 300 |
| No. validation shots | 80 | 80 | 100 | 100 |
| No. test shots | 120 | 120 | 200 | 200 |

## 3.1 Disruption Prediction on Existing Devices

The ample data can be obtained from existing tokamak devices. In this section, the predictors are trained by all slice samples from the training set in Table 2. The E-CAAD predictor is trained with normal samples from non-disruption training set and disruption precursor samples from disruption training set. The CAAD predictor is only trained with normal samples from non-disruption training set. In the model training, non-disruption samples with the same number of disruption precursor samples used by E-CAAD are added to the training set of CAAD to keep the same number of samples trained by the two predictors. And two predictors share the same validation and testing set on each device. First, we assess the ability of the CAAD and E-CAAD predictors in identifying disruption precursors. Figure 2 and Figure 3 illustrate the relationship between $RCE$ sequences returned by the trained predictors and diagnostic signals for disruption shots on J-TEXT and EAST. The $RCE$ sequences are normalized by dividing the $RCE$ average during the normal operation period of the predicted disruption shot. Additionally, the x-axis endpoint in each figure corresponds to $t_{CQ}$. In Figure 2 (a), the disruption shot 1056920 is caused by a locked mode. $Mirnov_p$ represents one channel from the poloidal Mirnov array on J-TEXT, and the frequency and amplitude variations of $Mirnov_p$ are displayed. As approaching to $t_{CQ}$, the amplitude of $Mirnov_p$ increases, the frequency decreases, and finally mode locking occurs. Both CAAD and E-CAAD predictors exhibit a rise in $RCE$ sequences coinciding with the abnormal appearance on $Mirnov_p$, with the $RCE$ increasing as it approaches $t_{CQ}$. This confirms the reliability of CAAD and E-CAAD predictors in identifying disruption precursors. Notably, during the precursor period, the normalized $RCE$s returned by E-CAAD surpass that of CAAD, indicating E-CAAD's superior ability in identifying disruption precursors. Consequently, when employing $RCE$ threshold rules for disruption prediction on the same disruption shot, the alarm time returned by the E-CAAD predictor precedes that of CAAD. The disruption shot 1057685 on J-TEXT in Figure 2(b) is caused by plasma vertical displacement due to the loss of horizontal field control, while the disruption shot 74742 on EAST in Figure 3(b) is caused by VDE. As the plasma undergoes abnormal vertical displacement, the $RCE$s returned by both predictors start to rise, with E-CAAD consistently exhibiting a larger increase in normalized $RCE$ compared to CAAD. This result demonstrates the good generalization ability of the two predictors between different devices. In Figure 3(a), the disruption shot 62425 is caused by impurities entering the plasma, resulting in increased edge radiation. As approaching to $t_{CQ}$, the XUV signal grows, and disruption occurs as plasma electron density drops to 0. Both predictors exhibit an upward trend in $RCE$ during the anomalous occurrence, with E-CAAD displaying a larger increase than CAAD. In conclusion, the testing of CAAD and E-CAAD predictors with disruption shots from multiple devices and various disruption types demonstrates their ability to distinguish disruption precursor samples from non-disruption samples. Moreover, E-CAAD exhibits a stronger ability in identifying disruption precursors compared to CAAD.

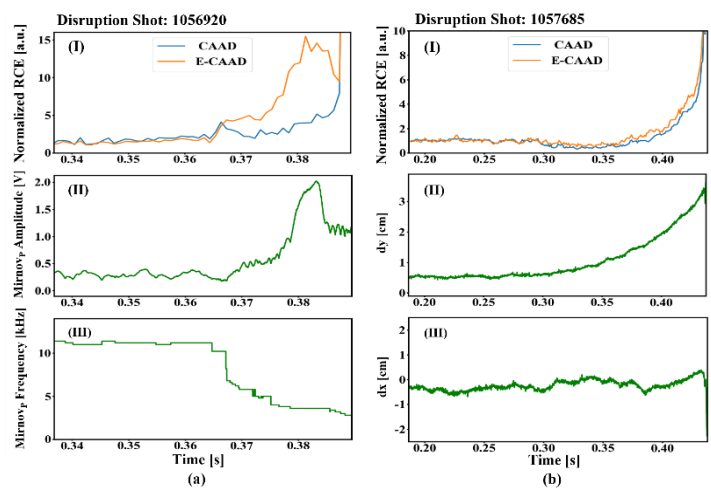

Figure 2 Comparison of the recognition ability of disruption precursors between CAAD and E-CAAD predictors on J-TEXT

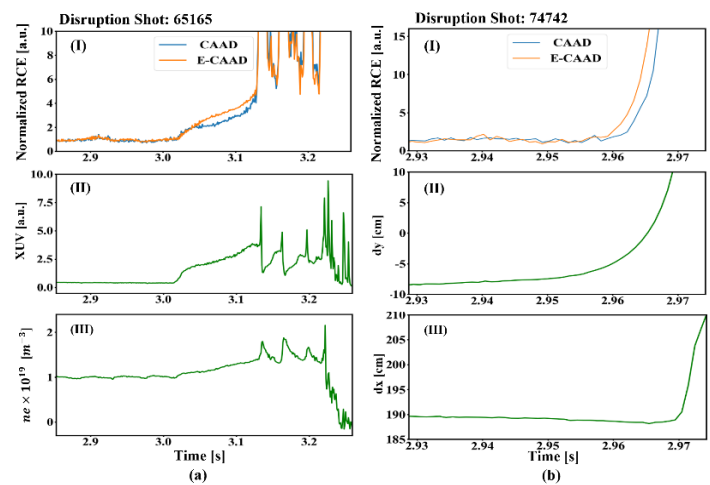

Figure 3 Comparison of the recognition ability of disruption precursors between CAAD and E-CAAD predictors on EAST

In disruption prediction, True positive (TP) refers to a successfully predicted disruption discharge. False positive (FP) refers to non-disruption discharges predicted as disruptions, also known as false alarms. True negative (TN) refers to non-disruption discharge predicted correctly. False negative (FN) refers to a disruption discharge not predicted, also known as the miss alarm and tardy alarm. The metrics used to evaluate the performance of disruption predictors often include accuracy, true positive rate (TPR), and false positive rate (FPR). They are calculated as follows:

$$Accuracy = \frac{TP + TN}{TP + FP + TN + FN} \quad (3\text{-}1)$$

$$TPR = \frac{TP}{TP + FN} \quad (3\text{-}2)$$

$$FPR = \frac{FP}{FP + TN} \quad (3\text{-}3)$$

The disruption predictor needs to reserve sufficient reaction time for the MGI system, which is referred to as the minimal warning time. And the alarms issued within the minimal warning time before plasma current quench time ($t_{CQ}$) are considered tardy alarms and are counted as false negatives. J-TEXT is a small-sized tokamak with a relatively shorter time scale for disruption to take place[43],[44]. Previous study found that the precursor durations of the disruption shots on J-TEXT is generally shorter than that on EAST[45]. Therefore, following the approach of previous studies, a minimal warning time of 10ms is selected for J-TEXT, while 20ms is chosen for EAST in this article. Figure 4 presents the receiver operating characteristic (ROC) curves of CAAD and E-CAAD predictors on J-TEXT and EAST. The area under curve (AUC) of the ROC quantifies the overall performance of the predictor, considering both true positive rate (TPR) and false positive rate (FPR). A higher AUC indicates a better overall performance of the predictor. Ideally, a perfect classifier would have an ROC AUC equal to 1. In Figure 4 (a), although there is a small portion on the ROC curve of CAAD in the upper left corner of E-CAAD on J-TEXT, the AUC of the improved predictor is greater than that of CAAD. This result indicates that the performance of the E-CAAD predictor is generally better than that of CAAD predictor, especially in terms of robustness. Figure 4 (b) shows the ROC curves of the two predictors on EAST, and the result still shows that E-CAAD predictor performs better than CAAD predictor. Through the hyperparameter search, two predictors with optimal parameters are obtained on J-TEXT and EAST, and their prediction performance is shown in Table 3.

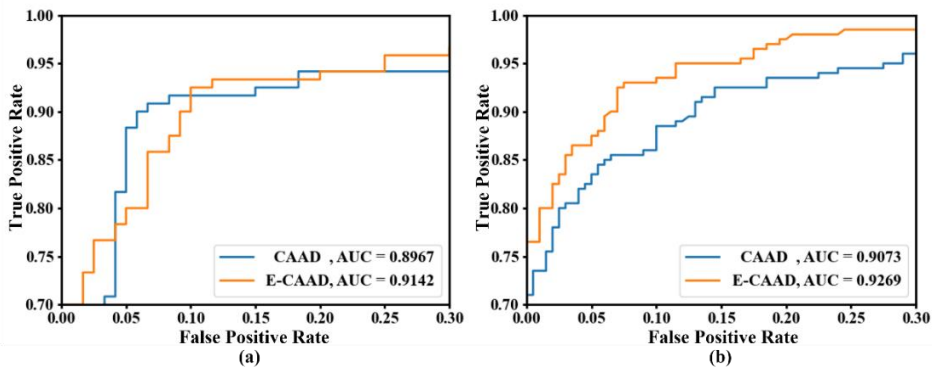

Figure 4 ROC curves of CAAD and E-CAAD predictors (a) on J-TEXT (b) on EAST. For E-CAAD predictor, $\alpha = 455$ on J-TEXT, $\alpha = 750$ on EAST.

Table 3 The prediction performance of CAAD and E-CAAD predictors on J-TEXT and EAST

| Device | Model | Accuracy [%] | TPR [%] | FPR [%] |
|---|---|---|---|---|
| JTEXT | CAAD | 93.16 | 90.83 | 6.67 |
|  | E-CAAD | 91.25 | 92.50 | 10.00 |
| EAST | CAAD | 89.75 | 84.50 | 5.00 |
|  | E-CAAD | 91.25 | 86.00 | 3.50 |

Analyze the warning time returned by each predictor on the test shot set. Figure 5 illustrates the accumulated percentage of disruption predicted versus warning time for CAAD and E-CAAD predictors on J-TEXT and EAST. The curve of E-CAAD predictor is slightly higher to the right than CAAD predictor on both J-TEXT and EAST. This result indicates that the alarm time inferred by E-CAAD predictor is generally earlier than CAAD predictor, which is consistent with the conclusion in Figure 2 and Figure 3. As the durations of discharges on J-TEXT are within 1s, the premature alarm threshold set on J-TEXT is 200ms. And the premature alarm rates of CAAD and E-CAAD are 3.67% and 3.60%. The durations of discharges on EAST are mostly 7s-30s, and some even exceed 100s, so the premature alarm threshold set on EAST is 500ms. The premature alarm rates of CAAD and E-CAAD are 4.73% and 4.65%, respectively.

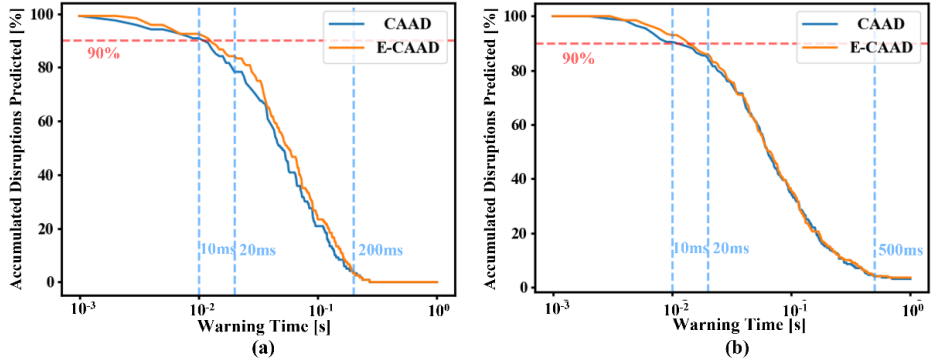

Figure 5 The accumulated percentage of disruption predicted versus warning time for CAAD and E-CAAD predictors (a) on J-TEXT (b) on EAST.

In summary, our findings indicate that with ample data training, E-CAAD predictor outperforms CAAD in recognizing disruption precursors. It exhibits superior disruption prediction performance, providing earlier alarm times for the same disruption shot.

### 3.2 Deployment from Scratch on New Devices

When a new tokamak starts operation, there will be shots available for disruption predictor training. As the device operates, useful shots are added to the dataset of the new device in chronological order. However, in the early stages of device operation, the dataset is scarce. And the proportion of disruption and non-disruption samples is unpredictable, which may cause imbalance. Deploying data-driven disruption predictors on the new device necessitates addressing the challenge of sample imbalance, with a desire for minimal reliance on new device data to achieve enough performance. The less new device data used means the faster the predictor is deployed on the new device. The CAAD and E-CAAD predictors address the issue of sample imbalance. In this section, we study the deployment speed of these two predictors on the new device. The deployment method starting from scratch involves initiating training with the first shot of the new device and regularly add the latest shots to the training set for predictor retraining. The evaluation compares the speed at which predictor performance improves as training data transition from scarcity to abundance.

In this section, deployment experiments from scratch are conducted using CAAD and E-CAAD predictors on both J-TEXT and EAST devices. The training set, which consists of disruption and non-disruption shots as mentioned in Section 3.1, is sorted in chronological order based on the experiment time. To simulate the temporal distribution of shots in the early stages of the new device's operation, a consecutive set of 400 shots is randomly selected for each device. The ratio and temporal distribution of non-disruption and disruption shots in the consecutive set are unpredictable before the deployment experiment. In Figure 6 and Figure 7, the background color shows the distribution of non-disruption shots and disruption shots on J-TEXT and EAST, respectively. The green portion represents the non-disruption shots, while the red portion represents disruptions. The predictor is trained starting from the first shot, and sequentially the 400 shots is added to the training set. After every new shot added, the model is retrained. This comparative experiment between CAAD and E-CAAD predictors keeps the same number of training samples for both predictors during each retraining iteration. For the E-CAAD predictor, only disruption precursor samples of the disruption shot are added to the training set when the disruption shot occurs; Conversely, during the normal operation period of this disruption shot, the same number of non-disruption samples as the disruption precursor samples are selected and added to the training set of the CAAD predictor. The performance evaluation of each retrained predictor is conducted on the test set mentioned in Section 3.1, and the corresponding accuracy, true positive rate (TPR), and false positive rate (FPR) were obtained.

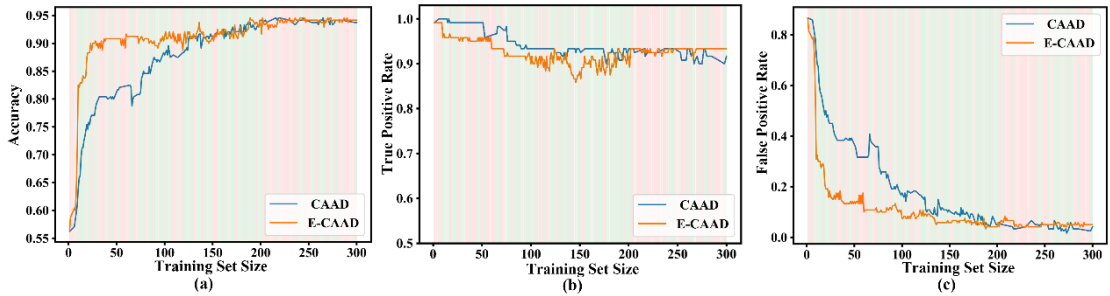

Figure 6 Deployment experiment of CAAD and E-CAAD predictors starting from scratch on J-TEXT

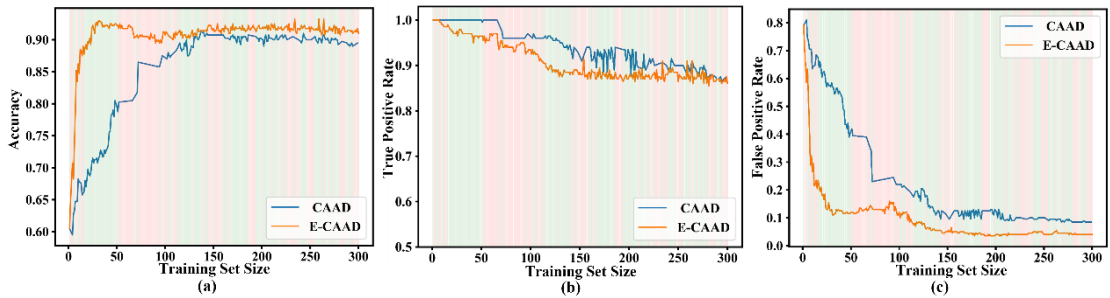

Figure 7 Deployment experiment of CAAD and E-CAAD predictors starting from scratch on EAST

Figure 6 shows the relationship between the performance of the CAAD and E-CAAD predictors and the size of training set on J-TEXT. The first shot trained is a normal shot, and at this location, the performance of the two predictors is the same. Starting from the second shot, the accuracy of E-CAAD has significantly increased compared to CAAD. This is because the second shot is a disruption shot, and E-CAAD trained with disruption precursor samples has a stronger recognition ability for disruption precursor samples than the CAAD trained with only non-disruption samples. During the initial stages when the training set is scarce, the E-CAAD predictor exhibits a more significant increase in accuracy compared to CAAD predictor as the training set size grows. This is evident in Figure 6(a), where E-CAAD predictor reaches a stable high-performance state earlier than CAAD predictor. This indicates that E-CAAD predictor achieves faster deployment in the experiment starting from scratch on J-TEXT. Furthermore, FPRs of these two predictors are shown in Figure 6(c). At the location of the first shot, FPRs are 86.66% for both predictors, indicating extremely poor performance. As the training set increases, FPRs decrease until they stabilize below 10%. During this process, FPR of the E-CAAD predictor decreases faster compared to the CAAD predictor. And TPRs of the predictors are shown in Figure 6(b). At the location of the first shot, the TPRs are close to 1. Research indicates that this is because the alarms for disruption shots include a large number of alarms issued during the normal operation period of the disruption shots. This type of alarm, similar in nature to false alarms for normal shots shown in Figure 6(b), is referred to as a premature alarm. As mentioned in Section 3.1, since the test set contains some disruption shots with precursor durations less than the minimum warning time of the MGI system, after training by ample data, both CAAD and E-CAAD predictors achieve TPRs of around 90%. Therefore, in Figure 6(b), as the size of training set increases, the TPRs start to decrease from around 1 until stabilizing at around 90%. In this process, compared to the CAAD predictor, the TPR of the E-CAAD predictor drops to a stable state more quickly, reflecting a faster decrease in the proportion of premature alarms in the TPR. These observations in Figure 6 highlight the E-CAAD predictor's ability to achieve faster deployment on J-TEXT. Figure 7 shows the relationship between the performance of CAAD and E-CAAD predictor and the size of training set on EAST. The E-CAAD predictor also demonstrates faster deployment compared to CAAD predictor in the experiment starting from scratch on EAST.

In summary, E-CAAD predictor can effectively use disruption precursor samples when the disruption shot occurs, enhancing the predictor's capability to identify disruption precursors. This results in a faster deployment speed of the predictor on new devices compared to the CAAD predictor.

## 4 Cross-tokamak Deployment Strategy Comparison

Due to differences in structure, operating parameters, control systems, or other factors between tokamaks, data-driven disruption predictors must use data from the new device to adapt to the operating environment of the new device when deployed across tokamaks[34]. Plasma disruption will cause serious economic losses on future new tokamaks. Therefore, it is hoped that the predictor will use as little data as possible from the new device to achieve high-performance disruption prediction as soon as possible on the new device. Current study on disruption prediction based on multi device data mixing[35] has indicated that the data from existing devices contains valuable physical information that can improve the disruption prediction performance of the predictor on the target device. This section focuses on leveraging data from existing devices to minimize predictor's reliance on new device data to achieve enough performance on new device. And future tokamak devices, such as ITER, have larger sizes and higher discharge performance compared to existing devices[34]. To simulate this situation, in this section, J-TEXT is regarded as the small-sized existing device, while EAST is regarded as the large-sized new device, and a strategy study is conducted using existing device data to assist predictors in deploying on the new device. Two deployment strategies are designed: Fine-tuning the model trained on existing devices and Mixing data from existing devices.

**Strategy 1: Fine-tuning the predictor trained on existing device**

Neural network-based disruption predictors typically begin with randomly initialized parameters before training, lacking any useful information. During training, these parameters are optimized by using a large dataset. After training on abundant data from the existing device, the network parameters of the predictor not only contain information about the specific operating environment of the existing device but also include valuable physical information for disruption prediction. On the other hand, in the early stages of new device operation, the dataset is scarce and insufficient for the model to learn the comprehensive physical information about all types of disruptions. Nevertheless, the dataset does contain information about the unique operating environment of the new device. Therefore, one approach is to use the network parameters of the pre-trained high-performance predictor from the existing device as the initial network parameters for the predictor on the new device. Then, fine-tuning can be performed using a small amount of data from the new device to adapt the predictor to its specific operating environment. In this section, the fine-tuning strategy uses the E-CAAD predictor for cross-tokamak deployment on EAST, starting from scratch. In each retraining iteration, the initial network parameters of the predictor are set to the network parameters of the high-performance predictor pre-trained on J-TEXT. This allows for the transfer of valuable knowledge from the existing device to the new device while enabling adaptation to the specific operating environment of EAST.

**Strategy 2: Mixing data from existing device**

One approach to cross-tokamak deployment of predictors is to train the predictor by mixing data from both the existing and new devices. In this section, we study a mixing-data based cross-tokamak deployment strategy to deploy E-CAAD predictor on EAST, starting from scratch. The strategy involves mixing the EAST data with the abundant data from J-TEXT as the training set during each retraining iteration. The initial network parameters for each retraining are randomly initialized. In this experiment, we use the training set of J-TEXT (525 shots) from Section 3.1 for mixing, while training on EAST starts from the first shot. The total number of shots in the training set remains constant at 526, with each retraining iteration adding one shot from EAST and removing one shot from J-TEXT until all the shots from J-TEXT are exhausted. However, in the initial stages of the new device's operation, there is limited data available, and the training set, which is a mixture of data from the new device and a large amount of data from the existing device, becomes extremely imbalanced. This imbalance can cause the neural network predictor to pay less attention to the data from the new device during training. To address this issue, further improvements are made to the loss function of E-CAAD predictor, allowing it to give more attention to EAST samples when the EAST dataset has a smaller proportion, and automatically adjust the attention weight with changes in the size of the EAST training set in each retraining iteration. The improved loss function is as follows:

$$l = \sum_{i=1}^{n}[y \cdot RCE_i^y + (1-y) \cdot \alpha \cdot RCE_i^{y-1}] \tag{4-1}$$

$$loss = \frac{1}{n_{J-TEXT} + n_{EAST}}[l_{J-TEXT} + \beta \cdot l_{EAST}], \quad \beta = \begin{cases} \frac{n_{J-TEXT}}{n_{EAST}}, & \frac{n_{EAST}}{n_{J-TEXT}} < 1 \\ 1, & \frac{n_{EAST}}{n_{J-TEXT}} \geq 1 \end{cases} \tag{4-2}$$

Where $l$ is the sum of the loss terms for all samples, $l_{J-TEXT}$ represents $l$ of all samples on J-TEXT, $l_{EAST}$ represents $l$ of all samples on EAST. $n_{J-TEXT}$ is the number of J-TEXT slice samples in training set. $n_{EAST}$ is the number of EAST slice samples in training set. $\beta$ is the weight that adjusts the attention of the predictor to EAST samples.

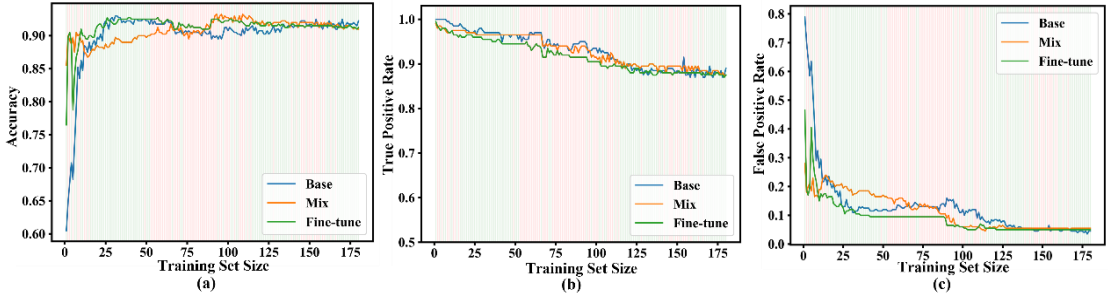

Figure 8 Comparison of cross-tokamak deployment strategies based on E-CAAD predictor.

The deployment strategy of E-CAAD predictor on EAST starting from scratch in section 3.2 is used as the base strategy. This base strategy does not use information from the existing device data, and the initial network parameters are randomized for each retraining. Two cross-tokamak deployment strategies are designed in this section, namely Fine-tuning the predictor trained on the existing device and mixing data from the existing device, are compared with the base strategy, and the comparative results are shown in Figure 8. The results demonstrate that during the period of data scarcity in the training set, the predictor using fine-tuning strategy is the fastest to achieve stable high performance in accuracy, TPR and FPR. This indicates that using existing device data can accelerate the deployment of the predictor on the new device. Among the compared cross-tokamak deployment strategies, the fine-tuning strategy enables the predictor to achieve the fastest deployment on the new device. For the mixing data deployment strategy, in the case where only the data from the first shot on EAST is used in the mixed training set, the performance of the predictor on the new device is an accuracy of 86%, TPR of 98.5%, and FPR of 27.5%, which is the best among all the compared strategies. However, during periods of data scarcity, as the size of the training set increases, the performance improvement of the predictor is the lowest among comparison strategies, and the accuracy and FPR curves reach a stable state at the latest. This suggests that even though the predictor pays more attention to the data of new device during the period of data scarcity, the large amount of mixed training data from the existing device misleads the predictor, thus slowing down its deployment on the new device. If multiple existing devices are used to assist in training the predictor in a mixed data strategy, it may make the predictor pay more attention to the common information related to plasma disruption in the training data, weaken the attention to the characteristic information of each existing device in the training data, and thus make the deployment speed of the predictor faster. Due to the current inability of the J-TEXT team to obtain trainable data from other devices than EAST, this strategy will be tested in the future.

In summary, this section confirms that the data from existing devices can accelerate the deployment of disruption predictor on new devices. And among the designed strategies, the fine-tuning strategy achieves the faster deployment on new devices; The advantage of the mixing data strategy is that it enables the predictor to achieve relatively good predictive performance right from the beginning of its deployment on the new device.

## 5  Summary

This study addresses the issue of data utilization in data-driven disruption predictor during cross tokamak deployment. In order to adapt to the environment where the proportion of positive and negative samples of the dataset on future tokamak are unbalanced is unpredictable, we propose an Enhanced Convolutional Autoencoder Anomaly Detection (E-CAAD) predictor. The E-CAAD predictor can be trained using only normal samples from non-disruption shots and can also be trained using disruption precursor samples when available. This approach not only overcomes the sample imbalance issue of supervised learning predictors but also overcomes the inefficient dataset utilization of traditional anomaly detection predictors. The network architecture of the E-CAAD predictor includes multiple layers of convolutional layers that extract the features from raw signal of each input diagnostic, enabling effective learning of the physical knowledge contained in the diagnostic signals. Furthermore, the performance comparisons are conducted between the E-CAAD predictor and the Convolutional Autoencoder Anomaly Detection (CAAD) predictor trained only by non-disruption samples on J-TEXT and EAST. With ample data training, E-CAAD predictor demonstrates outperforms CAAD in recognizing disruption precursors. It exhibits superior disruption prediction performance, providing earlier alarm times for the same disruption shot. In the experiment of deployment from scratch, E-CAAD predictor exhibits faster deployment on new devices compared to CAAD predictor. Finally, experiments conducted using J-TEXT as the existing device and EAST as the new device lead to the design of two strategies: Fine-tuning the predictor trained on existing device and Mixing data from the existing device. Comparisons between strategies reveal that the data from existing devices can assist in faster deployment of the predictor on the new device. Among the strategies, fine-tuning strategy enables the fastest deployment, while mixing data strategy achieves high predictive performance from the start of deployment. And the advantage of the mixing data strategy is that it enables the predictor to achieve relatively good predictive performance right from the beginning of its deployment on the new device.

In the future, the J-TEXT team will focus on two aspects of research. Firstly, we will continue to explore methods for the fast and high-performance deployment of disruption predictors on future devices. This will involve leveraging data from multiple existing devices, including approaches such as data fusion and extracting common physical features to facilitate cross-tokamak predictor deployment. Secondly, the anomaly detection model based on convolutional autoencoder discussed in this paper can provide information on the contribution of each input diagnostic signal and even each channel within the diagnostics. We plan to use this model to investigate the underlying physical mechanisms of plasma disruption. These studies will provide crucial theoretical support for future disruption prediction of the ITER reactor.

## Acknowledgments

The authors are very grateful for the help of J-TEXT team. This work is supported by the National MCF Energy R&D Program of China under Grant No. 2022YFE03040004 and by the National Natural Science Foundation of China under Grant No.51821005.

## References

[1] Schuller, F., Disruptions in tokamaks. Plasma Physics and Controlled Fusion, 1995. 37(11A): p. A135.
[2] Riccardo, V., Disruptions and disruption mitigation. Plasma physics and controlled fusion, 2003. 45(12A): p. A269.
[3] De Vries, P., et al., Survey of disruption causes at JET. Nuclear fusion, 2011. 51(5): p. 053018.
[4] Pautasso, G., et al., The ITER disruption mitigation trigger: developing its preliminary design. Nuclear Fusion, 2018. 58(3): p. 036011.
[5] Sias, G., et al., A locked mode indicator for disruption prediction on JET and ASDEX upgrade. Fusion Engineering and Design, 2019. 138: p. 254-266.
[6] Kates-Harbeck, J., A. Svyatkovskiy, and W. Tang, Predicting disruptive instabilities in controlled fusion plasmas through deep learning. Nature, 2019. 568(7753): p. 526-531.
[7] Wroblewski, D., G. Jahns, and J. Leuer, Tokamak disruption alarm based on a neural network model of the high-beta limit. Nuclear Fusion, 1997. 37(6): p. 725.
[8] Sengupta, A. and P. Ranjan, Prediction of density limit disruption boundaries from diagnostic signals using neural networks. Nuclear fusion, 2001. 41(5): p. 487.
[9] Windsor, C., et al., A cross-tokamak neural network disruption predictor for the

JET and ASDEX Upgrade tokamaks. Nuclear fusion, 2005. 45(5): p. 337.
[10] Yang, Z., et al., A disruption predictor based on a 1.5-dimensional convolutional neural network in HL-2A. Nuclear Fusion, 2019. 60(1): p. 016017.
[11] Cannas, B., et al., Support vector machines for disruption prediction and novelty detection at JET. Fusion engineering and design, 2007. 82(5-14): p. 1124-1130.
[12] Vega, J., et al., Results of the JET real-time disruption predictor in the ITER-like wall campaigns. Fusion Engineering and Design, 2013. 88(6-8): p. 1228-1231.
[13] Lopez, J.M., et al. Implementation of the Disruption Predictor APODIS in JET's Real-Time Network Using the MARTe Framework. IEEE Transactions on Nuclear Science, 2014, 61(2) p. 741-744.
[14] Tinguely, R., et al., An application of survival analysis to disruption prediction via Random Forests. Plasma Physics and Controlled Fusion, 2019. 61(9): p. 095009.
[15] Hu, W., et al., Real-time prediction of high-density EAST disruptions using random forest. Nuclear Fusion, 2021. 61(6): p. 066034.
[16] Zheng, W., et al., Disruption predictor based on neural network and anomaly detection on J-TEXT. Plasma Physics and Controlled Fusion, 2020. 62(4): p. 045012.
[17] Zhong, Y., et al., Disruption prediction and model analysis using LightGBM on J-TEXT and HL-2A. Plasma Physics and Controlled Fusion, 2021. 63(7): p. 075008.
[18] Montes, K.J., et al., Machine learning for disruption warnings on Alcator C-Mod, DIII-D, and EAST. Nuclear Fusion, 2019. 59(9): p. 096015.
[19] Guo, B.H., et al., Disruption prediction on EAST tokamak using a deep learning algorithm. Plasma Physics and Controlled Fusion, 2021. 63(11): p. 115007.
[20] Zhong, Y. et al. Disruption prediction and model analysis using LightGBM on J-TEXT and HL-2A. Plasma Physics and Controlled Fusion, 2021, 63(7), p.075008.
[21] Yang, Z., et al., In-depth research on the interpretable disruption predictor in HL-2A. Nuclear Fusion, 2021. 61(12): p. 126042.
[22] Cannas, B., et al., A prediction tool for real-time application in the disruption protection system at JET. Nuclear Fusion, 2007. 47(11): p. 1559.
[23] Cannas, B., et al., Disruption forecasting at JET using neural networks. Nuclear fusion, 2003. 44(1): p. 68.
[24] Vega, J., et al., Adaptive high learning rate probabilistic disruption predictors from scratch for the next generation of tokamaks. Nuclear Fusion, 2014. 54(12): p. 123001.
[25] Pautasso, G., et al., On-line prediction and mitigation of disruptions in ASDEX Upgrade. Nuclear Fusion, 2002. 42(1): p. 100.
[26] Cannas, B., et al., An adaptive real-time disruption predictor for ASDEX Upgrade. Nuclear Fusion, 2010. 50(7): p. 075004.
[27] Zhang, Y., et al., Prediction of disruptions on ASDEX Upgrade using discriminant analysis. Nuclear Fusion, 2011. 51(6): p. 063039.
[28] Rea, C. and R.S. Granetz, Exploratory machine learning studies for disruption prediction using large databases on DIII-D. Fusion Science and Technology, 2018. 74(1-2): p. 89-100.
[29] Rea, C., et al., Disruption prediction investigations using machine learning tools

on DIII-D and Alcator C-Mod. Plasma Physics and Controlled Fusion, 2018. 60(8): p. 084004.

[30] Yoshino, R., Neural-net disruption predictor in JT-60U. Nuclear fusion, 2003. 43(12): p. 1771.

[31] Yoshino, R., Neural-net predictor for beta limit disruptions in JT-60U. Nuclear fusion, 2005. 45(11): p. 1232.

[32] Jang, J. and S. Yoon, Feature concentration for supervised and semisupervised learning with unbalanced datasets in visual inspection. IEEE Transactions on Industrial Electronics, 2020. 68(8): p. 7620-7630.

[33] Nguyen, M.H., Impacts of unbalanced test data on the evaluation of classification methods. International Journal of Advanced Computer Science and Applications, 2019. 10(3).

[34] Wenninger, R., et al. Power handling and plasma protection aspects that affect the design of the DEMO divertor and first wall. in Proceedings of the of 26th IAEA Fusion Energy Conference, Kyoto, Japan. 2016.

[35] Windsor, C., et al., A cross-tokamak neural network disruption predictor for the JET and ASDEX Upgrade tokamaks. Nuclear fusion, 2005. 45(5): p. 337.

[36] Zhu, J., et al., Scenario adaptive disruption prediction study for next generation burning-plasma tokamaks. Nuclear Fusion, 2021. 61(11): p. 114005.

[37] Dormido-Canto, S., et al., Development of an efficient real-time disruption predictor from scratch on JET and implications for ITER. Nuclear Fusion, 2013. 53(11): p. 113001.

[38] Vega, J., et al., Adaptive high learning rate probabilistic disruption predictors from scratch for the next generation of tokamaks. Nuclear Fusion, 2014. 54(12): p. 123001.

[39] Murari, A., et al., Adaptive learning for disruption prediction in non-stationary conditions. Nuclear Fusion, 2019. 59(8): p. 086037.

[40] Goodfellow, I. et al. Deep learning (Vol. 1). Cambridge：MIT press，2016：Chapter 14, pp. 499-507.

[41] Hinton, G.E. and R. Zemel, Autoencoders, minimum description length and Helmholtz free energy. Advances in neural information processing systems, 1993. 6.

[42] Farrow, C.L., et al., Nyquist-Shannon sampling theorem applied to refinements of the atomic pair distribution function. Physical Review B, 2011. 84(13): p. 134105.

[43] Ding, Y.H., et al., Overview of the J-TEXT progress on RMP and disruption physics. Plasma Science and Technology, 2018. 20(12): p. 125101.

[44] Liang, Y., et al., Overview of the recent experimental research on the J-TEXT tokamak. Nuclear Fusion, 2019. 59(11): p. 112016.

[45] Ai, X.K., et al., Tokamak plasma disruption precursor onset time study based on semi-supervised anomaly detection. Nuclear Engineering and Technology, 2023.